\documentclass[
aps,
pra,
twocolumn,
superscriptaddress
]{revtex4-2}

\usepackage[utf8]{inputenc}
\usepackage{amsmath,amssymb,amsfonts}
\usepackage{graphicx}
\usepackage{bm}
\usepackage{braket}
\usepackage{xcolor}
\usepackage{hyperref}
\usepackage[misc]{ifsym}
\usepackage{float}

\begin{document}

\title{Thermal Quantum Correlations in Coupled Andreev Spin Qubits with Tunable Superconducting Phase and Spin–Orbit Interaction}

\author{Mohammad Reza Pourkarimi \textsuperscript{\Letter}}

\affiliation{Physics Department, Faculty of Basic Sciences, Salman Farsi University of Kazerun, Kazerun, Iran }

\author{Mina Shiri}
\affiliation{Faculty of Physics, Semnan University, P.O.Box 35195-363, Semnan, Iran}

\author{Mehdi Monemi}
\affiliation{Centre for Wireless Communications (CWC), University of Oulu, 90570 Oulu, Finland}
\affiliation{Department of Electrical and Computer Engineering, Salman Farsi University of Kazerun, Kazerun 7319673544, Iran}

\author{Vadim Ohanyan}
\affiliation{Laboratory of Theoretical Physics, Yerevan State University,  1 Alex Manoogian Str., 0025 Yerevan, Armenia}
\affiliation{CANDLE, Synchrotron Research Institute, 31 Acharyan Str., 0040 Yerevan, Armenia}

\begin{abstract}
We investigate thermal quantum correlations in a system of two coupled superconducting spin qubits described by an effective Andreev spin-qubit Hamiltonian in the presence of spin–orbit interaction. Using Local Quantum Fisher Information (LQFI) and Local Quantum Uncertainty (LQU) as quantum-correlation quantifiers, we analyze the effects of the superconducting phase difference, tunneling amplitude, spin–orbit coupling, and temperature on the nonclassical properties of the system. Analytical expressions for the thermal density matrix are obtained and employed to evaluate both quantities. Our results show that quantum correlations decrease monotonically with increasing temperature, while stronger tunneling and spin–orbit interaction significantly enhance their robustness. Moreover, the superconducting phase introduces a pronounced periodic behavior through the modulation of the effective exchange couplings, leading to constructive and destructive interference regimes that strongly influence the correlations. By analyzing the energy spectrum of the effective Hamiltonian, we demonstrate that the enhancement of quantum correlations is closely associated with an increased energy gap between the ground and first excited states, which suppresses thermal excitations and stabilizes the correlated ground state. Furthermore, a comparative analysis of LQFI, LQU, and negativity is presented under the same parameter conditions, providing further insight into their respective behaviors and the relationship between quantum correlations and entanglement. LQFI is consistently larger than or equal to LQU throughout the investigated parameter space. These findings reveal the microscopic mechanism governing thermal quantum correlations in Andreev spin qubits and highlight the important roles of phase engineering, spin–orbit interaction, and tunneling in protecting quantum resources in hybrid superconducting quantum devices.

\end{abstract}

\maketitle

\textsuperscript{\Letter} email: mrpourkarimy@gmail.com


\section{Introduction}
\label{intro}
The field of quantum computation has experienced rapid growth in recent decades \cite{Nielsen_Chuang_2010}. Significant advances in this area have led to the development of novel frameworks for addressing problems that are computationally intractable for classical computers \cite{Preskill_2018}. Quantum computers offer fundamental advantages, including ultra-secure quantum communication \cite{Bennett_2014}, efficient processing of large-scale data \cite{Biamonte_2017, Schuld_2014, 365700}, and the acceleration of search algorithms \cite{Grover_1998}. In this framework, solid-state platforms have been extensively investigated as promising candidates for quantum information processing \cite{Loss_1998}. Such systems are artificially engineered to simulate the dynamics of quantum particles. Among the available architectures, superconducting qubits \cite{Huang_2020, Kjaergaard_2020, devoret2004superconductingqubitsshortreview} and semiconductor-based spin qubits \cite{Harvey_2022, Loss_1998, Kloeffel_2013, Awschalom_2013, Khedif_2021, MOHAMED2022105693} stand out as leading contenders for scalable quantum computation \cite{Preskill_2018}. Although spin qubits generally benefit from longer coherence times \cite{Veldhorst_2014}, the realization of long-range coupling mechanisms and scalable architectures continues to pose substantial technical challenges.

Despite significant advances in various qubit architectures, the scalability of quantum computing remains one of the principal obstacles to the realization of large-scale quantum processors. In contrast, superconducting qubits, although characterized by comparatively shorter coherence times, offer important practical advantages due to the possibility of precise control of two-qubit gates and their compatibility with scalable circuit architectures \cite{Krinner_2019, POURKARIMI202327}.

In recent years, hybrid architectures combining the advantages of superconducting and spin qubits have attracted considerable attention as promising platforms for scalable quantum computation. Within this context, Andreev spin qubits \cite{Chtchelkatchev_2003, Spethmann_2022} have emerged as an innovative realization of hybrid superconducting-spin systems \cite{Sauls_2018, Hays_2021}. In such systems, a quantum dot is embedded within a Josephson junction \cite{Choi_2000}, allowing the charge and spin degrees of freedom to be engineered and controlled with high precision. Unlike natural atomic systems, quantum dots possess highly tunable physical properties, making them attractive for quantum-information applications. Furthermore, the coupling between qubits is mediated by the superconducting Josephson supercurrent, enabling effective long-range interactions and providing a natural route toward the implementation of quantum gates.

Quantum entanglement constitutes one of the most fundamental features of quantum mechanics and quantum information theory, with broad applications in quantum computation \cite{Shor1999}, quantum communication \cite{Nielsen_Chuang_2010, PRXQuantum.3.010311}, quantum sensing \cite{Giovannetti_2004}, quantum integrated sensing and communication \cite{11669037} and quantum key distribution \cite{Bennett_2014}. More recent studies have shown that quantum correlations beyond entanglement may also serve as valuable resources for quantum-information processing. One of the most prominent measures of such correlations is quantum discord \cite{Ollivier_2001}. This concept was independently introduced by Ollivier and Zurek \cite{Ollivier_2001}, and by Henderson and Vedral \cite{Henderson_2001, PhysRevLett.90.050401}, through the investigation of the discrepancy between two inequivalent quantum generalizations of classical mutual information. Their results demonstrated that even separable states may contain genuinely quantum correlations.

Although quantum discord is a powerful measure for identifying nonclassical correlations, its evaluation is generally difficult because it involves complicated optimization procedures, especially for arbitrary mixed states. To overcome these difficulties, Girolami \textit{et al.} introduced the local quantum uncertainty (LQU) as a discord-like measure based on the concept of skew information and quantum uncertainty \cite{PhysRevLett.110.240402}. Owing to its analytical tractability and the absence of optimization over measurements, LQU has been extensively investigated in a variety of quantum systems. Furthermore, Werlang \textit{et al.} \cite{PhysRevA.80.024103} demonstrated that, even under Markovian dynamics, entanglement may exhibit sudden death, whereas quantum discord remains nonzero and decays more smoothly.

LQU is also closely related to quantum Fisher information (QFI) \cite{bera2014rolequantumcorrelationmetrology, Kim_2018, Dhar_2015, Girolami_2014}, which plays a central role in quantum metrology and quantum estimation theory. QFI establishes an important conceptual bridge between quantum-information theory and skew information \cite{Wigner_1963}, and both quantities have been proposed as reliable indicators of nonclassical correlations in multipartite quantum systems. In particular, local quantum Fisher information (LQFI) and LQU are intrinsically connected through the notion of quantum uncertainty, and their interplay provides valuable insight into the structure and robustness of quantum correlations. Consequently, the behavior of LQU and LQFI has attracted considerable attention in recent years \cite{Muthuganesan_2021, Haseli_2020, Slaoui_2019, Mohamed_2022, Yurischev_2023, Elghaayda_2022, 2022EPJP..137..548B, El_Bakraoui_2022}.

Recently, Andreev spin qubits have emerged as promising candidates for quantum-information processing due to their unique combination of superconducting coherence and spin degrees of freedom. In particular, the effective Hamiltonian describing coupled superconducting spin qubits in the presence of spin--orbit interaction has attracted growing interest because it provides a versatile framework for engineering controllable quantum interactions, phase-dependent couplings, and nontrivial exchange mechanisms. Owing to the recent development of this platform, many fundamental quantum-information aspects of the model still remain unexplored.

Although several studies have investigated the electronic structure, transport properties, and coherent manipulation of Andreev spin qubits, the behavior of quantum correlations in such systems has not yet been systematically explored. To the best of our knowledge, quantum-information measures such as LQFI and LQU have not been investigated for coupled superconducting spin qubits described by the Andreev spin-qubit Hamiltonian. In particular, the combined influence of superconducting phase difference, spin--orbit interaction, tunneling amplitude, and thermal fluctuations on the robustness of quantum correlations remains largely unknown.

Furthermore, LQFI is directly related to quantum metrology and parameter estimation, while LQU is connected to the skew-information framework. Since Andreev spin qubits are promising candidates for future quantum technologies, these quantities provide information that is directly relevant to quantum sensing and quantum-information processing.

The present work aims to fill this gap by providing a comprehensive investigation of thermal quantum correlations in Andreev spin qubits through the simultaneous analysis of LQFI and LQU. Studying these two quantities together is particularly important because, although both are rooted in quantum uncertainty theory, they capture different aspects of nonclassical correlations and may exhibit qualitatively different responses to external parameters and thermal effects. Therefore, a comparative analysis of LQFI and LQU allows one to obtain a deeper understanding of the structure and stability of quantum correlations in hybrid superconducting-spin systems.
Furthermore, a comparative analysis of LQFI, LQU, and negativity as a measure of quantum entanglement is presented providing further insight into their respective behaviors and the relationship between quantum correlations and entanglement.

More specifically, we investigate the interplay between temperature, superconducting phase difference, tunneling strength, and spin--orbit interaction, and analyze how these parameters influence the generation, suppression, and protection of quantum correlations. Our results reveal rich phase-dependent behaviors, including constructive and destructive interference effects induced by the superconducting phase, as well as the nontrivial role of spin--orbit coupling in enhancing quantum correlations and improving their robustness against thermal fluctuations. These findings provide new insight into the quantum-information properties of Andreev spin qubits and may be relevant for the design of scalable hybrid quantum devices.\\

The rest of the article is structured as follows: Section \ref{section2} and Section \ref{section3}  provides a concise review of LQFI and LQU as quantifiers of quantum correlations. In Section \ref{section4}, the proposed Andreev spin-qubit model is introduced, and its corresponding thermal density matrix is derived. Finally, Section \ref{section5} presents the main findings and concluding remarks.\\

\section{Local Quantum Fisher Information}\label{section2}

Consider a quantum density operator $\rho_{\theta}$ as a function of an unknown sensing parameter $\theta$ represented as an ensemble of states:
\begin{equation}
\label{eq:hjkj3657}
    \rho_{\theta} = \sum_{i} p_{i(\theta)} \rvert {\psi}_{i(\theta)}\rangle\langle{\psi}_{i(\theta)}\rvert
\end{equation}
where $p_i(\theta)$ is the probability associated with the $i$-th pure state $\rvert {\psi}_{i(\theta)}\rangle$, satisfying the normalization condition  $\sum_{i} p_{i(\theta)}=1$. To extract information regarding $\theta$, a set of Positive Operator-Valued Measures (POVMs) $\mathcal{M}$ is employed, defined as $\mathcal{M} = \{M_x | M_x \geq 0, \sum_{x} M_{x} = \mathbb{I}\}$. In the simplest case of a standard projective measurement in the $Z$-basis, the set is $\mathcal{M} = \{M_0, M_1\}$, where $M_0 = \rvert{0}\rangle\langle{0}\rvert$ and $M_1 = \rvert{1}\rangle\langle{1}\rvert$. The statistical sensitivity attainable by a specific measurement $\mathcal{M}$ is quantified by the  Fisher Information (FI) \cite{paris2009quantum}:
\begin{equation}
    F(\rho_\theta|\mathcal{M}) = \sum_x \frac{1}{\mathrm{Tr}(M_x \rho_\theta)} \left[ \frac{\partial \mathrm{Tr}(M_x \rho_\theta)}{\partial \theta} \right]^2
\end{equation}
 The QFI is then defined by optimizing the FI over all possible measurements $\mathcal{M}$, formally represented as:
    \begin{equation}
        \mathcal{F}(\rho_\theta) = \sup_{\mathcal{M}} F(\rho_\theta|\mathcal{M})
    \end{equation}
    This can be calculated as
    \begin{align}
        \mathcal{F}(\rho_\theta)=\frac{1}{4} \mathrm{Tr}(\rho_\theta L^2_\theta)
    \end{align}
    where $L_\theta$ is the solution of the equation \\$\frac{\partial \rho_\theta}{\partial \theta}=\frac{1}{2} \left(L_\theta\rho_\theta + \rho_\theta L_\theta\right)$ \cite{braunstein1994statistical, paris2009quantum}.

    In many physical scenarios, the parameter-dependent state $\rho_\theta$ is generated by subjecting an  initial probe state, $\rho$, to a unitary evolution $U_\theta = e^{i H \theta}$ associated with a generator $H$. This results in the evolved state $\rho_\theta = U_\theta^\dagger \rho U_\theta$. The initial parameter-independent state can be expressed via its spectral decomposition as:
\begin{equation}
    \label{eq:rho5554}
    \rho = \sum_i p_i  \rvert{\psi_i}\rangle\langle{\psi_i}\rvert,
\end{equation}
where $\sum_i p_i = 1$. In contrast to the general case in \eqref{eq:hjkj3657}, the eigenvalues $\{p_i\}$ here are fixed and independent of the sensing parameter $\theta$. Under these conditions, the QFI becomes independent of $\theta$ and is determined solely by the initial state $\rho$ and the generator $H$ in the Hilbert space $\mathcal{H}$, expressed as \cite{Kim_2018}:
\begin{equation}
    \label{eq:hjkkxbdsa5g}
    \mathcal{F}(\rho, H) = \frac{1}{2} \sum_{\substack{i, j \\ p_i + p_j \neq 0}} \frac{(p_i - p_j)^2}{p_i + p_j} \left| \langle \psi_i | H | \psi_j \rangle \right|^2.
\end{equation}

    To characterize how a subsystem of a composite quantum system contributes to the total QFI and to quantify its internal non-classical correlations, the concept of LQFI is employed. While standard QFI measures the sensitivity of a state to an external parameter, LQFI is defined as the minimum information that can be extracted via local unitary operations, serving as a robust quantifier of quantum discord-type correlations \cite{Kim_2018}. Considering a composite Hilbert space $\mathcal{H} = \mathcal{H}_{A} \otimes \mathcal{H}_{B}$ corresponding to two subsystems $A$ and $B$, a bipartite state $\rho$ as a function of estimation $\theta$ subjected to a local unitary evolution on party $a$ can be represented as:
\begin{equation}
    \label{eq:4jfks9118}\rho_\theta = e^{-i \theta H_A \otimes \mathbb{I}_B} \rho e^{i \theta H_A \otimes \mathbb{I}_B}.
\end{equation}
The LQFI, denoted as $\mathcal{F}_\mathrm{L}(\rho)$, is defined as the minimum QFI over all possible local observables $H_A$ acting on subsystem $A$ \cite{Kim_2018}:
\begin{equation}
    \label{eq:dskfhsdkg03950439543}
    \mathcal{F}_\mathrm{L}(\rho) = \min_{H_A} \mathcal{F}(\rho, H_A \otimes \mathbb{I}_B)
\end{equation}

where $\mathcal{F}$ is defined as in \eqref{eq:hjkkxbdsa5g}. This minimization ensures that $\mathcal{F}_\mathrm{L}(\rho)$ captures the intrinsic quantumness of the state that cannot be eliminated by choosing a specific local measurement direction.
To ensure a physically consistent representation, the local observable $H_A$ is generally required to be a traceless normalized Hermitian operator. For the case of a qubit subsystem $A$, any such observable can be parameterized as $H_A = \vec{\sigma} \cdot \vec{r}$, where $\vec{r}$ is a unit vector ($|\vec{r}| = 1$) and $\vec{\sigma} = (\sigma_1,\sigma_2,\sigma_3)\equiv (\sigma_x, \sigma_y, \sigma_z)$ denotes the vector of Pauli matrices. This normalization ensures that $\text{Tr}(H_A^2) = d_A$, where $d_A = 2$ is the dimension of the qubit Hilbert space. Under these conditions, the LQFI for a bipartite state $\rho$ can be evaluated analytically as:

\begin{align}
        \mathcal{F}_\mathrm{L}(\rho) =1-\lambda_{\mathrm{max}} (M)
    \end{align}
    in which $\lambda_\mathrm{max}$ denotes the maximum eigenvalue and $M$ is a $3\times 3$ symmetric matrix with elements
    \begin{equation}
\begin{split}
        \label{lqfi}
        M_{lk} &= \sum_{\substack{i, j \\ p_i + p_j \neq 0}} \frac{2 p_i p_j}{p_i + p_j}
        \langle \psi_i | \sigma_l \otimes \mathbb{I}_B | \psi_j \rangle
        \langle \psi_j | \sigma_k \otimes \mathbb{I}_B | \psi_i \rangle,  \\&l,k\in\{1,2,3\}
\end{split}
   \end{equation}

\section{Local Quantum Uncertainty}\label{section3}
The total uncertainty relating to the measurement of an observable $H$ on quantum state $\rho$ is quantified as follows:
\begin{align}
     \mathrm{Var}(\rho, H)
    = \operatorname{Tr}(\rho H^{2})
    - \left( \operatorname{Tr}(\rho H) \right)^{2}.
\end{align}
As opposed to the classical uncertainty analysis wherein pure (completely known) state results in a zero uncertainty, here it is observed that $\mathrm{Var}(\rho, H)$ is not generally equal to zero if $\rho$ is a pure state. To investigate only the quantum part of the uncertainty, the {\it Wigner-Yanase quantum skew information (QSI)} \cite{Wigner_1963} is formulated as
\begin{align}
    \mathcal{I}(\rho, H) = -\frac{1}{2} \text{Tr}([\sqrt{\rho}, H]^2).
\end{align}
It can be shown that if commutator $[ \rho, H ] = 0$, then we have $\mathcal{I}(\rho, H ) = 0$ and all the uncertainty measured by $\mathrm{Var}(\rho, H)$ can be considered classical statistical noise. Considering a composite Hilbert space $\mathcal{H} = \mathcal{H}_{A} \otimes \mathcal{H}_{B}$ and a bipartite state $\rho$ represented in \eqref{eq:4jfks9118}, similar to the definition of LQFI in \eqref{eq:dskfhsdkg03950439543}, LQU  is defined as \cite{PhysRevLett.110.240402}
\begin{equation}
    \label{eq:dskfhsdkg0334543ff543}\mathcal{U}_{\mathrm{L}}(\rho) = \min_{H_A} \mathcal{I}(\rho, H_A \otimes \mathbb{I}_B)
\end{equation}
For the case of a qubit subsystem $A$ with dimension $d_A=2$, it can be shown that
    \begin{align}\label{lqu}
        \mathcal{U}_{\mathrm{L}}(\rho) =1-\lambda_{\mathrm{max}} (W)
    \end{align}
    where $W$ is a $3\times 3$ matrix with entries obtained by 
 \begin{align}       
        \omega_{ij} \equiv \mathrm{Tr}\!\left\{
        \sqrt{\rho}\, (\sigma_i \otimes \mathbb{I}_B)\,
        \sqrt{\rho}\, (\sigma_j \otimes \mathbb{I}_B)
        \right\}, \ \ i,j\in\{1,2,3\}
    \end{align}

    where $(\sigma_1,\sigma_2,\sigma_3)\equiv(\sigma_x,\sigma_y,\sigma_z)$. Considering the spectral decomposition of $\rho$ as in \eqref{eq:rho5554} using a complete orthogonal basis  $\rvert \psi_{i}\rangle$, leveraging the completeness property $\sum_{i}  \rvert \psi_{i}\rangle\langle \psi_{i} \rvert=\mathbb{I}$, it can be verified that the QSI is obtained as
    \begin{align}
        \mathcal{I}(\rho, H)
        =
        \frac{1}{2}
        \sum_{i,j}
        \left(\sqrt{p_i} - \sqrt{p_j}\right)^2
        \left|\langle \psi_i \mid H \mid \psi_j \rangle\right|^2.
    \end{align}
    The upper and lower bounds of QFI and LQFI can be expressed in terms of QSI and LQU respectively, as follows \cite{Slaoui_2019}:

  \begin{align}
    \label{eq:hfff1}
        \mathcal{I}(\rho, H) \leq \mathcal{F} (\rho, H) \leq 2\mathcal{I}(\rho, H),
        \\
        \label{inequality}
        \mathcal{U}_{\mathrm{L}}(\rho) \leq \mathcal{F}_{\mathrm{L}} (\rho) \leq 2 \mathcal{U}_{\mathrm{L}}(\rho),
    \end{align}

 It is directly inferred from  \eqref{inequality} that the local uncertainty of an state $\rho$ is upper bounded by LQFI.

Although LQFI and LQU both quantify nonclassical quantum
correlations, they are based on different physical concepts. LQFI is
closely related to the sensitivity of a quantum state to local
parameter variations and therefore has a direct connection with
quantum statistical distinguishability and quantum metrology. In
contrast, LQU is based on the quantum uncertainty associated with
local measurements. These complementary aspects have been explored
in different two-qubit systems, including double quantum dots with
tunneling asymmetry and two-qubit Heisenberg models with
Dzyaloshinskii--Moriya interactions~\cite{Moqine_2023, Moqine_2025}.

\section{The models}\label{section4}

\begin{figure*}[t] \centering \includegraphics[width=150mm]{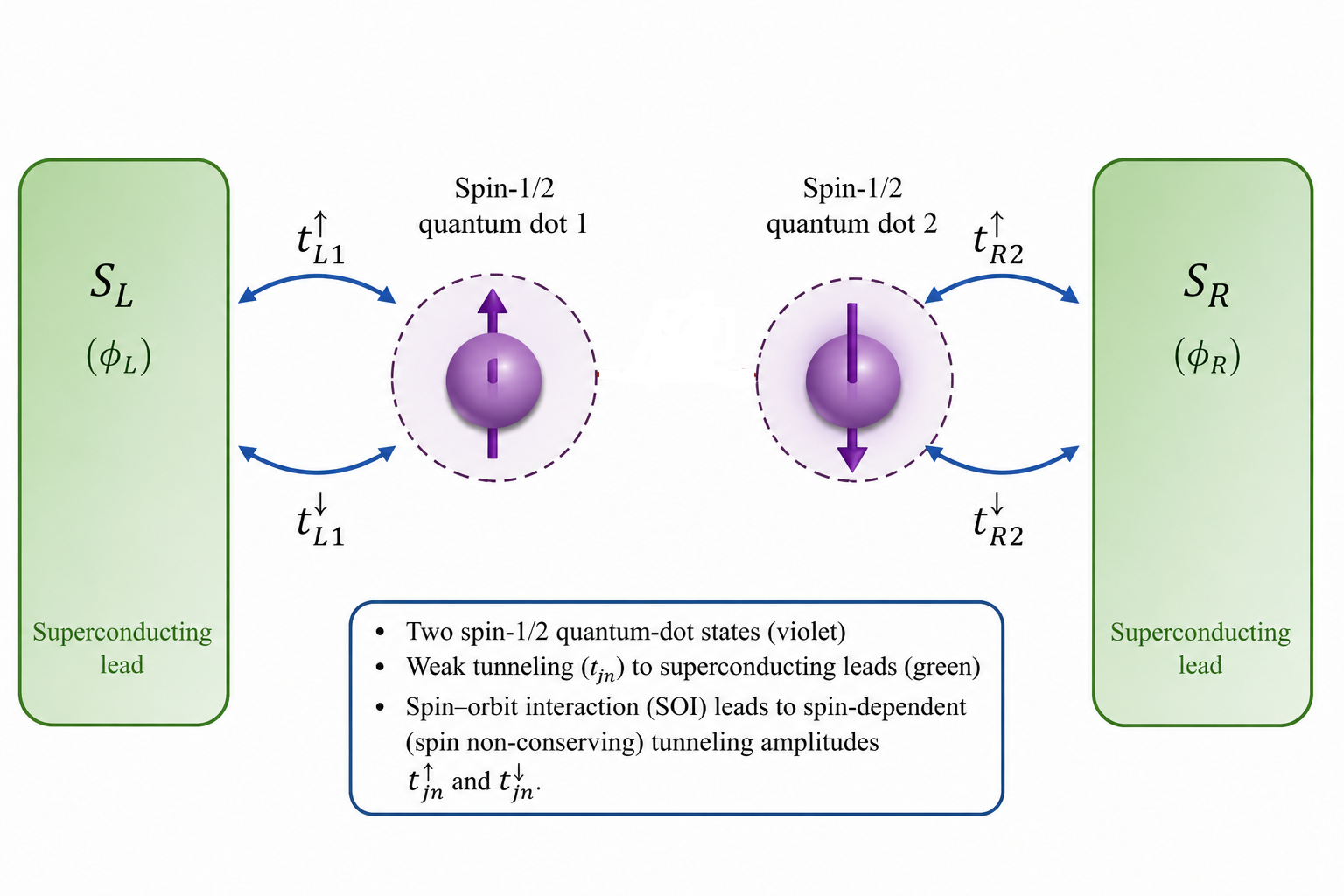} \caption{
Schematic illustration of two coupled superconducting spin qubits formed by spin-$1/2$ quantum dots connected to superconducting leads $S_L$ and $S_R$ with phases $\phi_L$ and $\phi_R$. The tunneling amplitudes between the quantum dots and the superconducting leads are denoted by $t_{jn}$. Due to the presence of spin--orbit interaction (SOI), the tunneling processes become spin dependent and may involve spin-flip transitions. The superconducting phase difference $\varphi=\phi_L-\phi_R$ controls the effective interaction between the qubits and plays a central role in the generation of quantum correlations.
} \label{fig:shem} \end{figure*}

The model of superconducting spin qubits consists of two quantum dots in two nanowires coupled to two superconducting leads by tunneling as displayed in figure \ref{fig:shem}.

By assuming large one-site Coulomb repulsion, negative dot tunneling from the Fermi energy and weak tunneling, we get a Hamiltonian with three components \cite{Spethmann_2022}.

The first one describes the  superconductor leads as follows
\begin{equation}\label{H1}
H_{1} = \sum_{jks}\varepsilon_{k}\,c^{\dagger}_{jks}c_{jks}
- \sum_{jk}\Big(\Delta_{j}c^{\dagger}_{jk\uparrow}c^{\dagger}_{j,-k\downarrow} + \text{H.c.}\Big),
\end{equation}
where $\varepsilon_{k}$ and $k$ are the single-particle energy and wave vector of spin $s \in \{\uparrow, \downarrow\}$, respectively.
$c^{\dagger}_{jks}$ and $c_{jks}$ are the creation and annihilation operators in the lead $j \in \{L,R\}$.
The parameter $\Delta_j = \Delta e^{-i\varphi_j}$ is defined in terms of the superconducting gap $\Delta$ and the superconducting phase $\varphi_j$, with the phase difference between the superconductors given by $\varphi = \phi_{L} - \phi_{R}$.

The dot Hamiltonian $H_{2}$ is expressed as
\begin{equation}\label{H2}
H_{2} = \sum_{ns}\epsilon\, d^{\dagger}_{ns}d_{ns}
+ C_{d}\, d^{\dagger}_{n\uparrow}d_{n\uparrow}d^{\dagger}_{n\downarrow}d_{n\downarrow},
\end{equation}
where $d^{\dagger}_{ns}$ and $d_{ns}$ are the creation and annihilation operators for an electron on the dot $n \in \{1,2\}$.
Here, $\epsilon$ denotes the dot energy level and $C_{d}$ represents the Coulomb repulsion.
We assume $\epsilon < 0$ and $C_{d}$ is sufficiently large such that each dot is singly occupied.

To derive the effective interaction between two superconducting spin qubits, we begin with the microscopic tunneling Hamiltonian describing the coupling between the quantum dots and the superconducting leads. In the presence of spin--orbit interaction (SOI), tunneling processes can rotate the spin of an electron. Consequently, the tunneling amplitudes acquire a matrix structure in spin space, which can be represented by unitary SU(2) matrices. The tunneling Hamiltonian is written as

\begin{equation}\label{H3}
H_{3} = \sum_{jnk} t_{jn}\,c^{\dagger}_{jk} U_{jn} d_{n} + \text{H.c.},
\end{equation}

where \(d_n=(d_{n\uparrow},d_{n\downarrow})^T\) and \(c_{j\mathbf{k}}=(c_{j\mathbf{k}\uparrow},c_{j\mathbf{k}\downarrow})^T\) are the electron spinors in dot \(n\) and superconducting lead \(j\), respectively. Here, \(t_{jn}\) denotes the spin-independent tunneling amplitude, while \(U_{jn}\) describes the spin rotation associated with tunneling between the corresponding dot and lead. Thus, the microscopic effect of SOI is encoded in the spin-dependent structure of the tunneling processes.

The effective spin interaction is obtained by eliminating the intermediate charge and quasiparticle degrees of freedom and retaining the low-energy dynamics of the two dot spins. In the regime considered here, the superconducting gap is large compared with the relevant dot energy scales, and tunneling is treated perturbatively. Virtual tunneling processes then generate an effective exchange interaction between the spins, even though the dots are not directly coupled by an ordinary magnetic exchange term. The resulting interaction depends on the tunneling amplitudes, the dot energy detuning, the superconducting phase difference \(\varphi\), and the spin rotations induced by SOI.

A convenient step in this derivation is to perform a unitary transformation of the electron spinors. By appropriately choosing the spin basis, the spin rotations associated with individual tunneling links can be absorbed into the definitions of the transformed operators. The remaining spin-dependent effect is characterized by a single effective rotation,

\begin{equation}
U=U_{R1}^{\dagger}U_{R2}U_{L2}^{\dagger}U_{L1}
=e^{i\alpha\mathbf{u}\cdot\mathbf{S}_1},
\end{equation}

where \(\alpha\) is the effective spin-rotation angle, \(\mathbf{u}\) is the unit vector specifying the SOI axis, and \(\mathbf{S}_1\) is the spin operator of the first qubit. We denote the spin operator of the second qubit in the transformed basis by \(\widetilde{\mathbf{S}}_2=U_{R}^{\dagger}{\mathbf{S}}_2U_{R}\). This basis transformation does not remove the physical effects of SOI; rather, it transfers them from the individual tunneling amplitudes into the effective interaction between the two spins.

Following the perturbative elimination of the intermediate degrees of freedom, the effective Hamiltonian can be expressed as

\begin{equation}\begin{split}\label{Hef}
H={}&J\mathbf{S}_1\cdot\widetilde{\mathbf{S}}_2\\
&+J\left(
e^{-i\alpha\mathbf{u}\cdot\mathbf{S}_1}
\mathbf{S}_1
e^{i\alpha\mathbf{u}\cdot\mathbf{S}_1}
\right)\cdot\widetilde{\mathbf{S}}_2\\
&+2J\cos\varphi\left[
\cos\frac{\alpha}{2}\,
\mathbf{S}_1\cdot\widetilde{\mathbf{S}}_2
-\sin\frac{\alpha}{2}\,
\mathbf{u}\cdot
(\mathbf{S}_1\times\widetilde{\mathbf{S}}_2)
\right]\\
&+J\sin\varphi\sin\frac{\alpha}{2}\,
\mathbf{u}\cdot
(\mathbf{S}_1-\widetilde{\mathbf{S}}_2).
\end{split}\end{equation}

Here, \(J\) is the effective exchange-energy scale determined by the microscopic tunneling and dot-energy parameters; in the symmetric setup considered in the reference model, it is estimated as \(J\simeq\Gamma^2/|\epsilon|\), with \(\Gamma_L=\Gamma_R=\Gamma\) and \(\Delta\gg|\epsilon|\). The parameter \(\varphi\) is the superconducting phase difference, whereas \(\alpha\) characterizes the net spin rotation due to SOI. 

To obtain a form that is more convenient for analyzing the spin correlations, we choose the SOI axis to lie along the \(z\) direction, namely, $\mathbf{u}=\mathbf{z}$. The spin operators can then be decomposed into components parallel and perpendicular to this axis: $\mathbf{S}_{1,\parallel}=S_{1z}\hat{\mathbf{z}}$, $\widetilde{\mathbf{S}}_{2,\parallel}=\widetilde S_{2z}\hat{\mathbf{z}}$, $\mathbf{S}_{1,\perp}=(S_{1x},S_{1y},0),$ and $\widetilde{\mathbf{S}}_{2,\perp}=(\widetilde S_{2x},\widetilde S_{2y},0).$
With this choice, the spin rotation acts about the \(z\) axis. The interaction can consequently be separated into longitudinal exchange, transverse exchange, antisymmetric exchange, and a phase-dependent field term. In the present work, we set \(U_R=I\), so that the rotated spin operator of the second qubit reduces to its unrotated form, \(\widetilde{\mathbf S}_2=\mathbf S_2\). This simplification eliminates the explicit basis rotation associated with \(U_R\), but it does not eliminate the effects of spin-orbit interaction (SOI) from the effective Hamiltonian. In particular, the parameter \(\alpha\) remains in the coupling coefficients \(h\), \(J_{\parallel}\), \(J_{\perp}\), and \(J_{\mathrm{DM}}\), thereby retaining the SOI-induced anisotropy, Dzyaloshinskii--Moriya (DM) interaction, and staggered-field contribution. Consequently, the resulting Hamiltonian is generally not reduced to an isotropic Heisenberg model.  Collecting the coefficients of these contributions gives the effective Hamiltonian

\begin{equation}\begin{split}\label{H}
H &= h(S_{1}^z - S_{2}^z)
+ J_{z} S_{1}^z S_{2}^z
+ J_{x}(S_{1}^x S_{2}^x + S_{1}^y S_{2}^y)\\
  &\quad
+ J_{DM}(S_{1}^x S_{2}^y - S_{1}^y S_{2}^x),
\end{split}\end{equation}
with
\begin{equation}\label{params}
\begin{aligned}
h &= J \sin\varphi \,\sin\!\frac{\alpha}{2}, \\[6pt]
J_{z} &= 2J\!\left(1 + \cos\varphi \,\cos\!\frac{\alpha}{2}\right), \\[6pt]
J_{x} &= 2J \cos\!\frac{\alpha}{2}\,\Big(\cos\varphi + \cos\!\frac{\alpha}{2}\Big), \\[6pt]
J_{\mathrm{DM}} &= -2J \sin\!\frac{\alpha}{2}\,\Big(\cos\varphi + \cos\!\frac{\alpha}{2}\Big),
\end{aligned}
\end{equation}
In the notation of Ref. \cite{Spethmann_2022}, the microscopic tunneling amplitudes are combined into quantities such as $\Gamma_j=\pi\rho_F t_{j1}t_{j2}$  where $\rho_F$ is the normal density of states per spin of the leads at the Fermi energy. For the symmetric case, \(\Gamma_L=\Gamma_R=\Gamma\), the effective exchange scale has the perturbative dependence $J\simeq \Gamma^{2}/\left| \epsilon \right|,$ apart from convention-dependent prefactors. Hence, \(J\) has dimensions of energy, while its magnitude is ultimately determined by the microscopic tunneling amplitudes, the density of states, and the relevant virtual excitation energy.

{
In the system considered here, the relevant low-energy states are associated with Andreev bound states localized in the superconducting weak links. In the regime where each relevant Andreev state is singly occupied, its spin degree of freedom can be represented by an effective spin-\(1/2\) operator. The two effective spins therefore describe the spin degrees of freedom associated with the two coupled superconducting spin qubits, rather than independent quasiparticle excitations that are retained explicitly in the effective model.

The reduction to a spin-only description relies on a separation of energy scales. Charge fluctuations and quasiparticle excitations are treated as virtual intermediate states and are integrated out when deriving the effective Hamiltonian. This approximation is justified when the relevant low-energy spin states are well separated from charge-excited configurations and real quasiparticle excitations, so that the latter do not participate directly in the low-energy dynamics. Their virtual effects are nevertheless retained through the effective exchange interactions and the spin–orbit- and phase-dependent coupling coefficients.
A detailed derivation of the Hamiltonian (\ref{H}), together with the underlying physical assumptions and the definitions of the relevant interaction parameters, can be found in Ref.~\cite{Spethmann_2022}.

The obtained effective Hamiltonian is the particular case of more general spin 1/2 $XYZ$ dimer Hamiltonian which has been considered earlier in the context of magneto-thermal and entanglement effects of non-conserving magnetization \cite{ada20, oha15, tor16},
\begin{equation}\begin{split}\label{eq: Ham_dimer}
H&=-h\left(g_1 S_1^z+g_2 S_2^z\right)+\\&
J\left\{\left(1+\gamma\right)S_1^xS_2^x+\left(1-\gamma\right)S_1^yS_2^y+\Delta S_1^z S_2^z\right\}\\&
+D\left(S_1^xS_2^y-S_1^yS_2^x\right).
\end{split}\end{equation}
Here the non-conserved magnetization originates from the non-homogeneous $g$-factors of the spin in the dimer and $XY$ anisotropy of symmetric exchange:

\begin{equation}
\begin{split}
\left [S^{z},H \right]&=2iJ\gamma \left ( S_{1}^{x}S_{2}^{y}+S_{1}^{y}S_{2}^{x} \right ), \\
\left [M^{z},H\right]&= i\left\{
\left(J\gamma g_{+}-D g_{-}\right)
\left(S_1^xS_2^y+S_1^yS_2^x\right)
\right. \\
&\left.
- Jg_-
\left(S_1^xS_2^y-S_1^yS_2^x\right)
\right\}, \\
g_{\pm}&=g_1\pm g_2, \
\end{split}
\end{equation}
 where $S^z=S_1^z+S_2^z$ is the $z$-component of the total spin, and $M^z=g_1S_1^z+g_2S_2^z$ is the magnetization. Due to non-homogeneous $g$-factors the magnetization operator is not proportional to the total $S^z$. Therefore, even for the vanishing $XY$ anisotropy, $\gamma=0$, when $S^z$ becomes a good quantum number, magnetization can be made conserved only when $g_2=g_1$. Non-conserved magnetization yields a series of exotic features for finite spin clusters \cite{oha15, tor16, ada20, ada24, ada24a, ada24b,bel14,tor18, oha12} as well as for the spin chains \cite{var19,kro20, pan20, bre20, jap21, bar23,oha20, yin24a, yin24b}. These features include such phenomena as non-linear dependence of the spectrum on magnetic field and monotonous dependence of the zero temperature magnetization on magnetic field within the single ground states \cite{oha15,bel14,ada24,oha12}, zero-temperature spin textures with ordered and disordered subsystems for the $g$-factors of different signs \cite{yin24a, yin24b, tor18}, unreachable saturation magnetization for the system with $XY$ anisotropy \cite{oha20,var19, pan20} and enhancement of the bipartite entanglement of the spin ground states \cite{ada20, ada24, ada24a, ada24b}. In the Ref. \cite{ada20} it was shown that for the spin dimer with the Hamiltonian (\ref{eq: Ham_dimer}) special values of the non-uniform $g$-factors can bring to the enhancement of the entanglement properties. Interestingly, the effective quantum-dot Hamiltonian obtained in the present paper can be obtained from Eq. (\ref{eq: Ham_dimer}) by setting $\gamma=0$, $g_2=-g_1=-1$, $J=J_x$, $J\Delta=J_z$ and $D=J_{DM}$. Interestingly, as was shown in Ref. \cite{ada20} this particular configuration of the $g$-factors brings to the maximum entanglement of the $\gamma\neq 0$ ground state
 \begin{eqnarray}
 &&\left|\Psi_{\pm}\right\rangle=\frac{1}{\sqrt{1+A_{\pm}^2}}\left(\left|\right\uparrow\uparrow\rangle+A_{\pm}\left|\right\downarrow\downarrow\rangle\right),\\
 &&A_{\pm}=\frac{hg_+ \pm\sqrt{h^2g_+^2+J^2\gamma^2+D^2}}{\sqrt{J^2+D^2}}. \nonumber
 \end{eqnarray}
 However, our Hamiltonian has a principal difference, absence of $XY$ anisotropy and conservation of the total $S^z$.

\begin{figure*}[t] \centering \includegraphics[width=150mm]{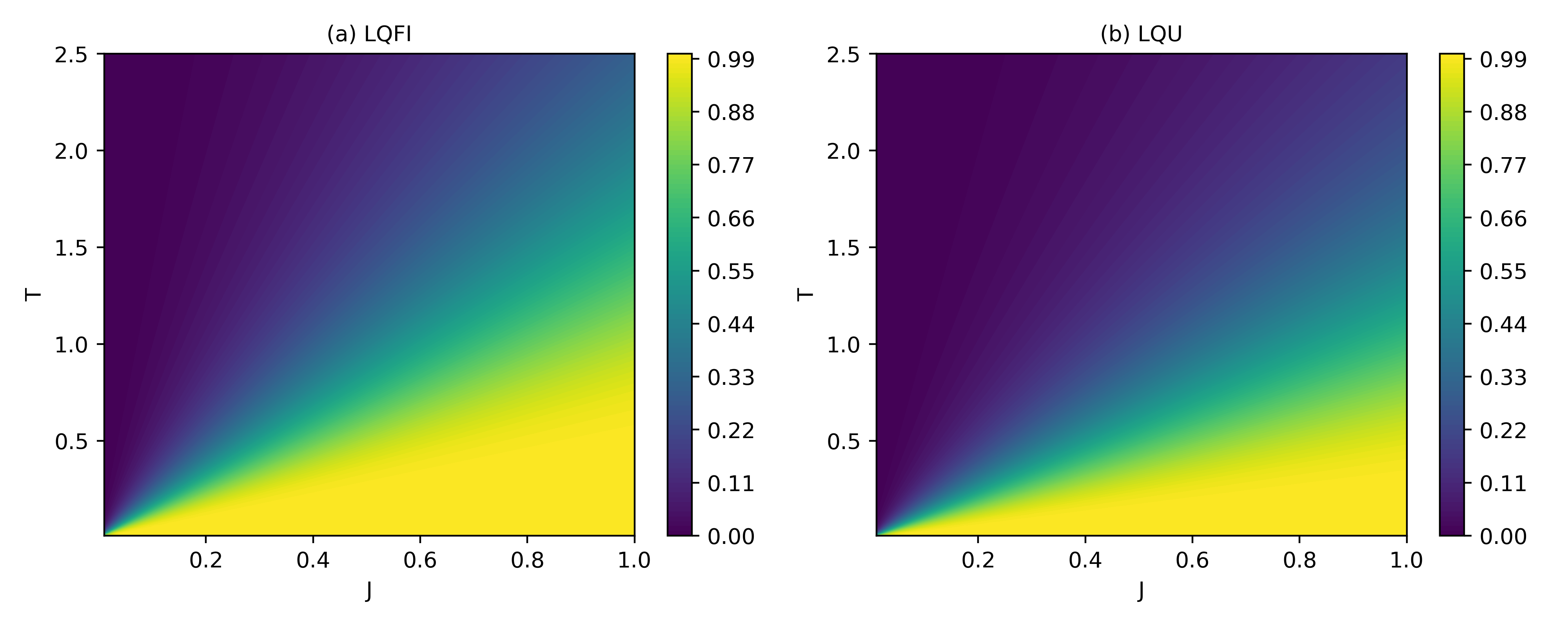} \caption{(a) LQFI and (b) LQU versus $T$ and $J$ for $\alpha=\varphi=\pi/8$.} \label{T_J} \end{figure*}

 \begin{figure*}[t] \centering \includegraphics[width=150mm]{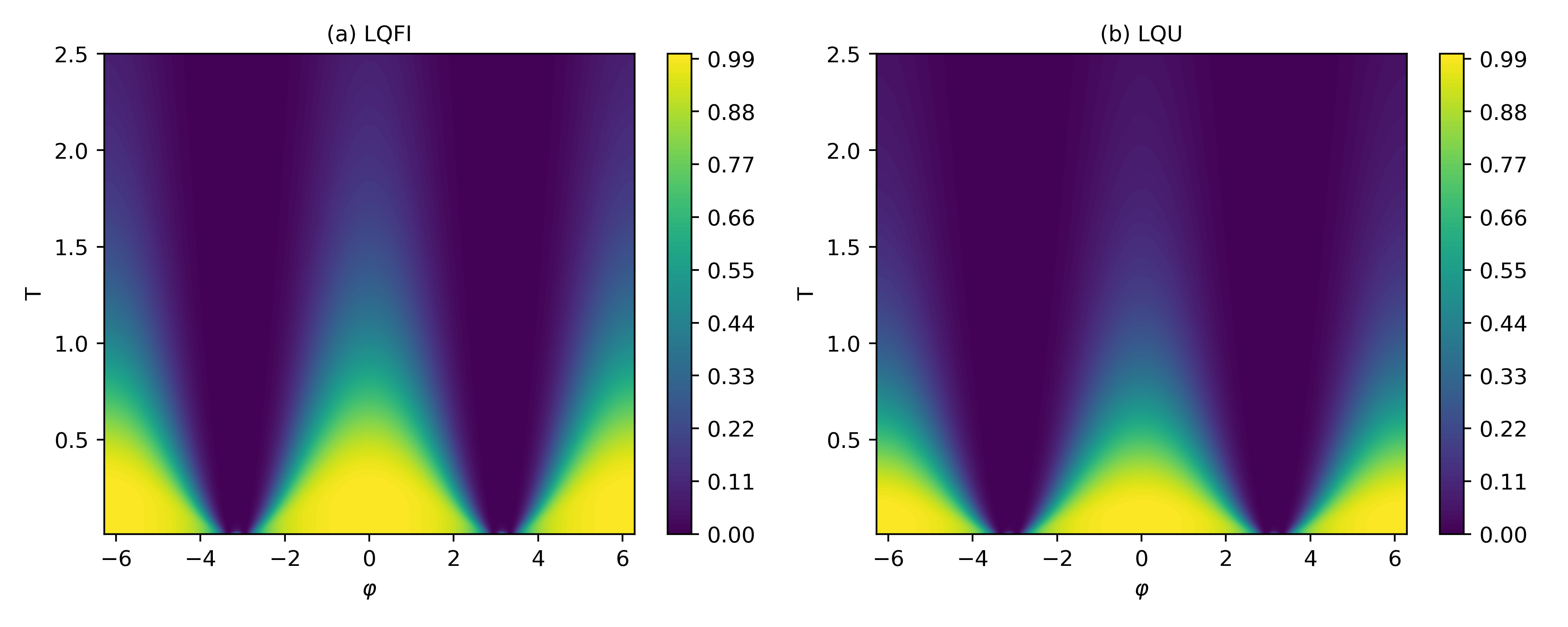} \caption{(a) LQFI and (b) LQU versus $\varphi$ and $T$ for $\alpha=\pi/8$ and $J=0.5$.} \label{T_phi} \end{figure*}

 \begin{figure*}[t] \centering \includegraphics[width=150mm]{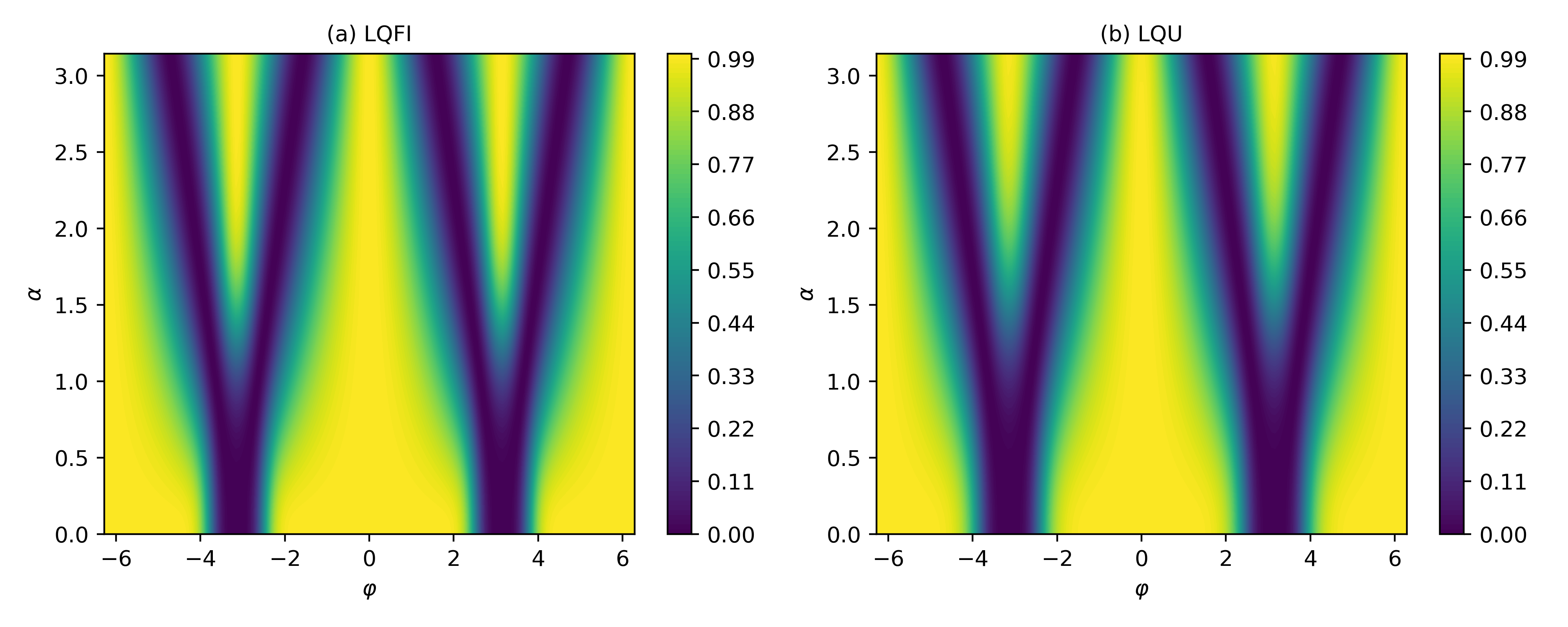} \caption{(a) LQFI and (b) LQU versus $\varphi$ and $\alpha$ for $J=0.5$ and $T=0.1$.} \label{al_phi} \end{figure*}

 \begin{figure*}[t] \centering \includegraphics[width=150mm]{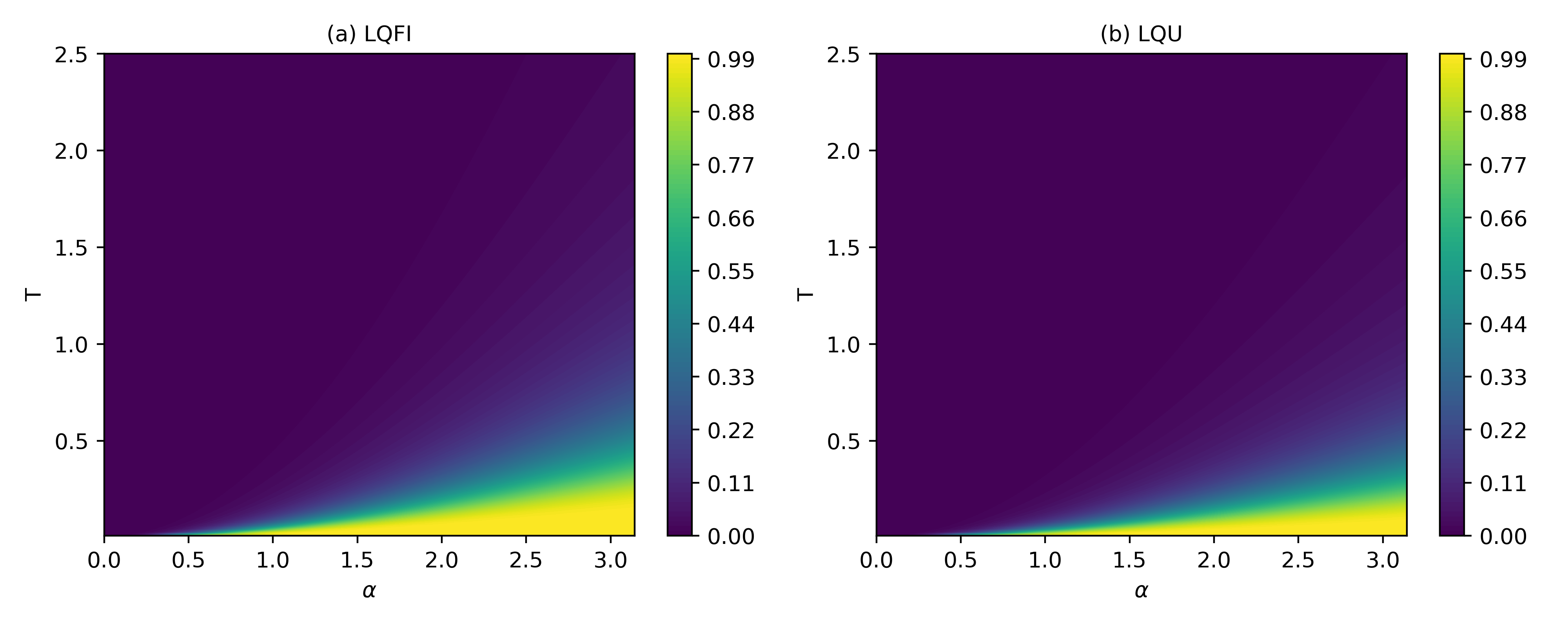} \caption{(a) LQFI and (b) LQU versus $\alpha$ and $T$ for $\varphi=\pi$ and $J=0.5$.} \label{T_al} \end{figure*}

 \begin{figure*}[t] \centering \includegraphics[width=150mm]{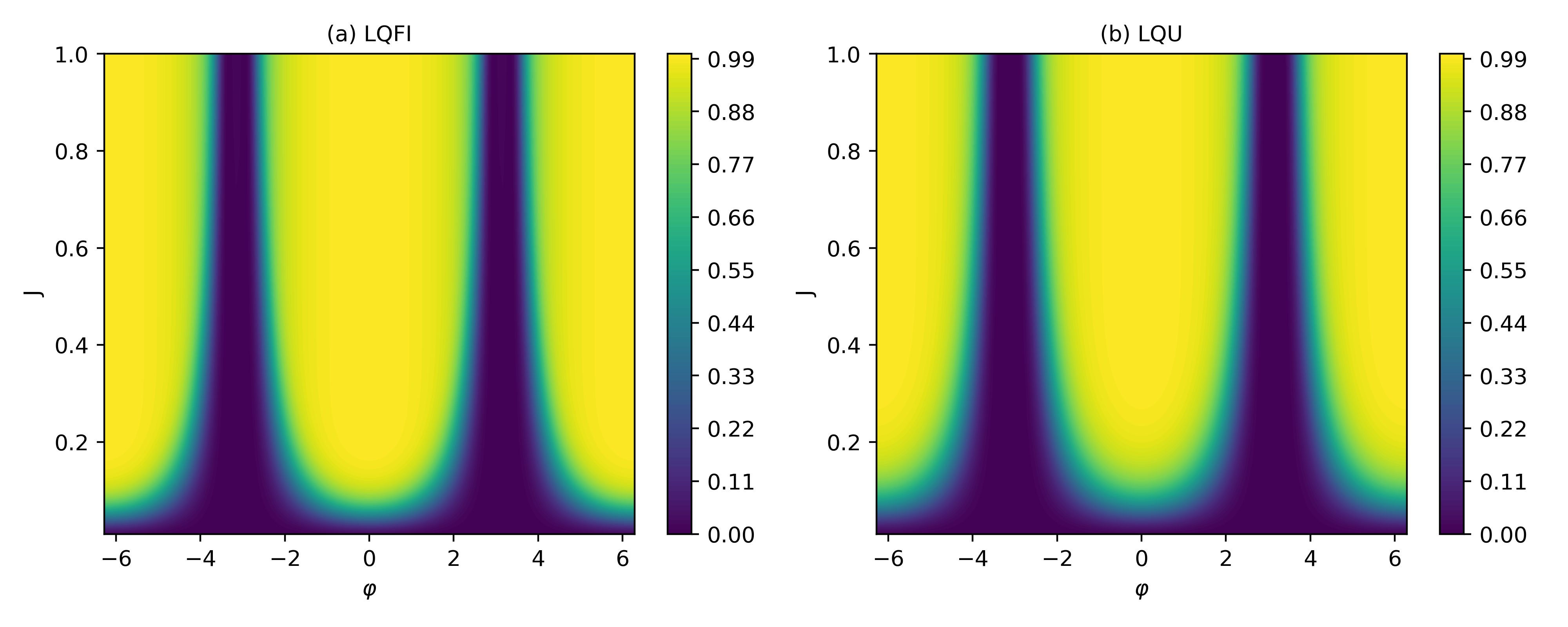} \caption{(a) LQFI and (b) LQU versus $\varphi$ and $J$ for $\alpha=\pi/8$ and $T=0.1$.} \label{J_phi} \end{figure*}

 \begin{figure*}[t] \centering \includegraphics[width=150mm]{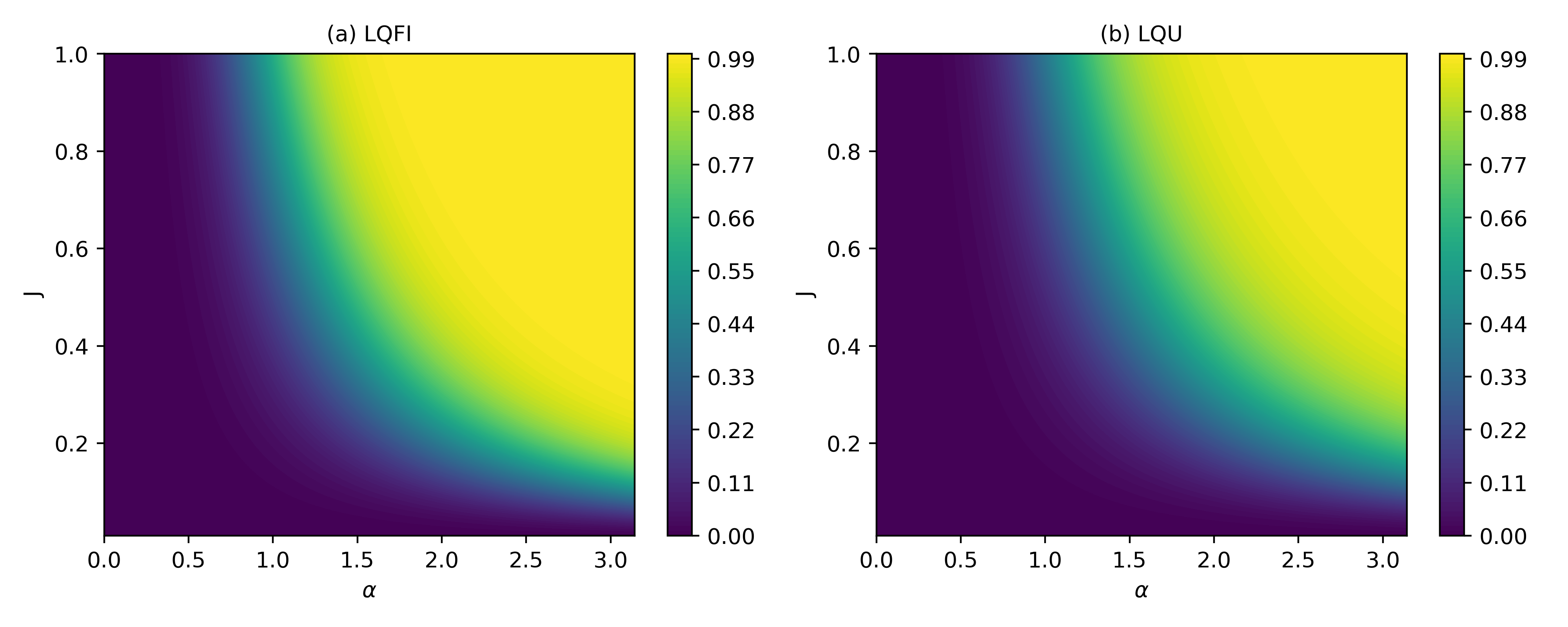} \caption{(a) LQFI and (b) LQU versus $\alpha$ and $J$ for $\varphi=\pi$ and $T=0.1$.} \label{J_al} \end{figure*}

\vspace{1cm}

In the standard basis, the Hamiltonian \eqref{H} takes the matrix form
\begin{equation}
\label{Hmatrix}
H = \begin{pmatrix}
 \tfrac{J_z}{4} & 0 & 0 & 0 \\
 0 & h - \tfrac{J_z}{4} & \tfrac{J_x}{2} + \tfrac{i J_{DM}}{2} & 0 \\
 0 & \tfrac{J_x}{2} - \tfrac{i J_{DM}}{2} & -h - \tfrac{J_z}{4} & 0 \\
 0 & 0 & 0 & \tfrac{J_z}{4}
\end{pmatrix}
\end{equation}

whose eigenvalues and eigenstates are
\begin{equation}
\begin{aligned}
E_{0} =-3J_z/4&&E_{1} = E_{2}=E_{3}=J_z/4
\end{aligned}
\end{equation}

\begin{equation}
\begin{aligned}
\ket{\psi_{0}}=&\frac{1}{\sqrt{1+\left| \left( \frac{-b}{a^{+}} \right) \right|^{2}}}(\frac{b}{a^{+}}\ket{01}+\ket{10})\\
\ket{\psi_{1}}
=&\ket{00}\\
\ket{\psi_{2}}=&\frac{1}{\sqrt{1+\left| \left( \frac{a^{-}}{b} \right) \right|^{2}}}(-\frac{a^{-}}{b}\ket{01}+\ket{10})\\
\ket{\psi_{3}}=&\ket{11}
\end{aligned}
\end{equation}
where $a^{\pm }=\frac{J_{x}}{2J}\left( \frac{e^{\frac{\pm i\alpha}{2}}}{Cos\left( \frac{\alpha}{2} \right)} \right)$ and $b=\frac{1}{J}\left( \frac{J_{z}}{2}-h \right)$.
\\
At thermal equilibrium at temperature $T$, the density operator of the two coupled superconducting spin qubits corresponding to the Hamiltonian \eqref{Hmatrix} is defined as
\begin{equation}\label{Trho}
\rho = \frac{1}{Z}\, e^{-\beta H},
\end{equation}
 where  $Z=Tr\left[exp(-\beta H) \right]$  is the partition function of the system and  $\beta = 1/k_{B}T$.
Here we set $k_{B}=1$ for simplicity. Based on Eqs.~\eqref{Hmatrix} and \eqref{Trho}, the density matrix takes the $X$-form:
\begin{equation}
\label{rhoab}
\rho_{AB} = \frac{1}{Z}
\begin{pmatrix}
 \nu & 0 & 0 & 0 \\
 0 & \eta^{+} & \gamma & 0 \\
 0 & \gamma^{*} & \eta^{-} & 0 \\
 0 & 0 & 0 & \nu
\end{pmatrix}.
\end{equation}
where  $Z=2 e^{-\frac{J_z}{4T}} \left(e^{\frac{J_z}{2T}} \cosh \left(\frac{\chi}{2 T}\right)+1\right)$, \\ $ \nu = e^{-\frac{J_z}{4T}}$,
$\eta^{\pm}=\frac{e^{\frac{J_z}{4T}} \left(\chi \cosh \left(\frac{\chi}{2 T^3}\right) \mp 2 h T^2 \sinh \left(\frac{\chi}{2 T^3}\right)\right)}{\chi}$ \\ and $\gamma=\frac{-T^2 ({J_x} + i J_{DM}) e^{-\frac{J_z}{4T}} \sinh \left(\frac{\chi}{2 T^3}\right)}{\chi}$  with\\ $\chi=\sqrt{4 h^2+J_{DM}^2+J_x^2}$.

\section{Results and discussions}\label{section5}

\subsection{Parameter Dependence of LQFI and LQU}
By substituting the density matrix (\ref{rhoab}) into Eqs.~(\ref{lqfi}) and (\ref{lqu}), we obtain the analytical expressions of LQFI and LQU for the two superconducting spin qubits coupled to a thermal bath. In the following, these quantities are employed to investigate the behavior of quantum correlations under the combined effects of temperature, superconducting phase difference, tunneling amplitude, and spin--orbit interaction.

Before discussing the numerical results, it is useful to emphasize the hierarchy relation between the considered quantities. It has been shown that LQU and LQFI are connected through inequalities originating from skew-information formalism and quantum estimation theory. In particular, LQU provides a lower bound for certain forms of LQFI, since LQFI captures a broader class of quantum fluctuations associated with local parameter estimation as shown in (\ref{inequality}). Therefore, from Eq. (\ref{inequality}) one generally expects $\mathrm{LQFI} \geq \mathrm{LQU}$, which is fully consistent with the results obtained in the present work. Throughout the parameter regimes investigated here, the numerical values of LQFI are found to be larger than those of LQU. This observation reflects the different definitions and normalizations of these two measures and should not be taken as a direct indication that LQFI is a more powerful or more robust measure than LQU. Instead, LQFI and LQU capture complementary aspects of quantum correlations. For a comprehensive review of discord-like measures and their different physical interpretations, see Ref. \cite{HU20181}.\\

Figure~\ref{T_J} presents contour plots of quantum correlations, quantified by LQFI and LQU, as functions of temperature $T$ and coupling strength $J$ for a fixed superconducting phase $\varphi=\pi/8$ and $\alpha=\pi/8$. The color scale represents the magnitude of the corresponding quantum-correlation measure, where brighter regions correspond to stronger correlations. It is clearly observed that both LQFI and LQU decrease monotonically with increasing temperature due to thermal fluctuations, which increase the mixedness of the density matrix and suppress quantum coherence. In contrast, increasing the coupling strength $J$ significantly enhances quantum correlations. Since $J$ is proportional to the tunneling amplitude, stronger tunneling increases the effective exchange interaction between the qubits and consequently promotes nonclassical correlations.

Moreover, LQFI remains systematically larger than LQU throughout the entire parameter region. Physically, this behavior can be understood from the different nature of the two measures. LQFI quantifies the maximal achievable precision in local quantum parameter estimation and is therefore highly sensitive to quantum fluctuations and local phase variations. In contrast, LQU measures the minimum quantum uncertainty associated with local observables. Because the optimization procedures involved in the definitions of LQFI and LQU are fundamentally different, LQFI captures a broader range of quantum statistical fluctuations and thus preserves larger values across the whole parameter space.

Figure~\ref{T_phi} illustrates the dependence of quantum correlations on the superconducting phase difference $\varphi$ and temperature $T$ for fixed values $\alpha=\pi/8$ and $J=0.5$. A pronounced periodic structure is observed with respect to $\varphi$, reflecting the interference nature of the phase-dependent terms in the effective Hamiltonian. The correlations attain their maximal values around $\varphi=0,\pm2\pi$, while they are strongly suppressed near $\varphi=\pm\pi$.

This behavior can be understood directly from the effective coupling coefficients given in Eq.~(\ref{params}). For $\varphi=0$, one has $\cos\varphi=1$ and $\sin\varphi=0$, and hence:

\begin{equation}\label{params2}
\begin{aligned}
J_z &= 2J\left(1+\cos\frac{\alpha}{2}\right), \\[6pt]
J_x &= 2J\cos\frac{\alpha}{2}
\left(1+\cos\frac{\alpha}{2}\right), \\[6pt]
J_{DM} &= -2J\sin\frac{\alpha}{2}
\left(1+\cos\frac{\alpha}{2}\right).
\end{aligned}
\end{equation}
Where the effective staggered magnetic field vanishes, $h=0$. In this regime, all coupling contributions reinforce each other constructively, leading to strong effective spin--spin interactions. Physically, the exchange interaction between the two qubits becomes maximal, allowing more efficient transfer of quantum information and stronger collective quantum behavior. Consequently, the system exhibits enhanced quantum coherence and stronger nonclassical correlations, explaining the maxima observed around $\varphi=0$.

In contrast, for $\varphi=\pi$, one has $\cos\varphi=-1$ and $\sin\varphi=0$, leading to
\begin{equation}\label{params3}
\begin{aligned}
J_z =2J\left(1-\cos\frac{\alpha}{2}\right), \\[6pt]
J_x =2J\cos\frac{\alpha}{2}
\left(-1+\cos\frac{\alpha}{2}\right).
\end{aligned}
\end{equation}
For small and moderate values of $\alpha$, the factor $\cos(\alpha/2)\approx1$, implying that both $J_z$ and $J_x$ become strongly suppressed. Consequently, the effective exchange interaction between the qubits weakens considerably, reducing entangling processes and suppressing nonclassical correlations. Therefore, the minima around $\varphi=\pm\pi$ originate from destructive interference between different virtual tunneling paths, which effectively reduces the coupling strengths in the Hamiltonian.
Increasing temperature gradually suppresses the phase-dependent structures due to thermal effects and enhanced mixedness of the density matrix. Nevertheless, the periodic dependence on $\varphi$ remains visible even at moderate temperatures, demonstrating the important role of superconducting phase interference in controlling quantum correlations.

Figure~\ref{al_phi} shows quantum correlations as functions of $\varphi$ and the spin--orbit interaction parameter $\alpha$ for fixed $J=0.5$ and low temperature $T=0.1$. It is evident that increasing $\alpha$ enhances quantum correlations, indicating the constructive role of spin--orbit interaction in generating nonclassical correlations. In particular, larger values of $\alpha$ strengthen the Dzyaloshinskii--Moriya interaction term $J_{DM}\propto\sin\frac{\alpha}{2},$ which introduces additional quantum fluctuations and enhances the coupling between spin components. Although the correlations are minimized near $\varphi=\pm\pi$, sufficiently large values of $\alpha$ partially compensate for this suppression and preserve finite correlations even in these regions.

Figure~\ref{T_al} presents contour plots of LQFI and LQU as functions of $\alpha$ and temperature $T$ for $\varphi=\pi$ and $J=0.5$. Since the phase is fixed at a value corresponding to destructive interference, the effective exchange couplings are strongly reduced. Consequently, quantum correlations are generally suppressed, especially for small values of $\alpha$. However, increasing the spin--orbit interaction strength substantially enhances the correlations and improves their robustness against thermal effects. This behavior demonstrates that spin--orbit interaction acts as an additional source of quantum coherence capable of partially counteracting thermal decoherence.

Figure~\ref{J_phi} illustrates the variation of quantum correlations as functions of $\varphi$ and $J$ for $\alpha=\pi/8$ and $T=0.1$. The periodic dependence on $\varphi$ is again clearly visible, with minimum values occurring around $\varphi=\pm\pi$. Increasing $J$ enhances the correlations because larger tunneling amplitudes strengthen the effective coupling between qubits. Nevertheless, even large values of $J$ cannot completely remove the suppression induced by the superconducting phase near $\varphi=\pm\pi$, indicating that phase-induced destructive interference dominates over tunneling-induced enhancement in these parameter regions.

Finally, Fig.~\ref{J_al} illustrates the dependence of quantum correlations on $J$ and $\alpha$ for fixed $\varphi=\pi$ and $T=0.1$. Both parameters positively contribute to the generation of quantum correlations. In particular, the combined effect of strong tunneling and strong spin--orbit interaction leads to maximal correlations. However, because the superconducting phase is fixed at the destructive-interference point $\varphi=\pi$, the correlations remain weaker than those obtained in the constructive regime around $\varphi=0$.

Furthermore, the spin--orbit interaction parameter $\alpha$ plays a crucial role through the trigonometric factors $\cos(\alpha/2)$ and $\sin(\alpha/2)$. For small $\alpha$, the Dzyaloshinskii--Moriya interaction is negligible. However, as $\alpha$ increases, the DM interaction becomes increasingly important and contributes significantly to the enhancement of quantum correlations, as observed in Figs.~\ref{al_phi},\ref{T_al} and \ref{J_al}.

Additionally, the effective staggered magnetic-field term $h \propto \sin\varphi \sin\frac{\alpha}{2}$
vanishes at $\varphi=0$ and $\varphi=\pi$, but becomes maximal near $\varphi=\pm\pi/2$. Consequently, intermediate regions of the phase diagrams may exhibit partially enhanced quantum correlations even when the exchange couplings are not maximal.

Our results also exhibit several similarities with previous investigations of Heisenberg spin systems and semiconductor spin-qubit models. In conventional Heisenberg-type systems, quantum correlations are generally enhanced by stronger exchange couplings and suppressed by thermal fluctuations, which is consistent with the behavior observed here. However, the present Andreev spin-qubit model possesses additional phase-dependent interference mechanisms and spin--orbit-induced couplings that are absent in ordinary Heisenberg models. In particular, the interplay between superconducting phase difference and spin--orbit interaction leads to richer correlation structures and greater tunability compared with standard spin-chain systems. Moreover, the persistent hierarchy relation $\mathrm{LQFI}\geq\mathrm{LQU}$ observed throughout the parameter space is consistent with previous studies of quantum-correlation measures in condensed-matter systems, while the phase-controlled suppression and revival of correlations constitute distinctive features of the present hybrid superconducting-spin architecture.

\begin{figure*}[t] \centering \includegraphics[width=150mm]{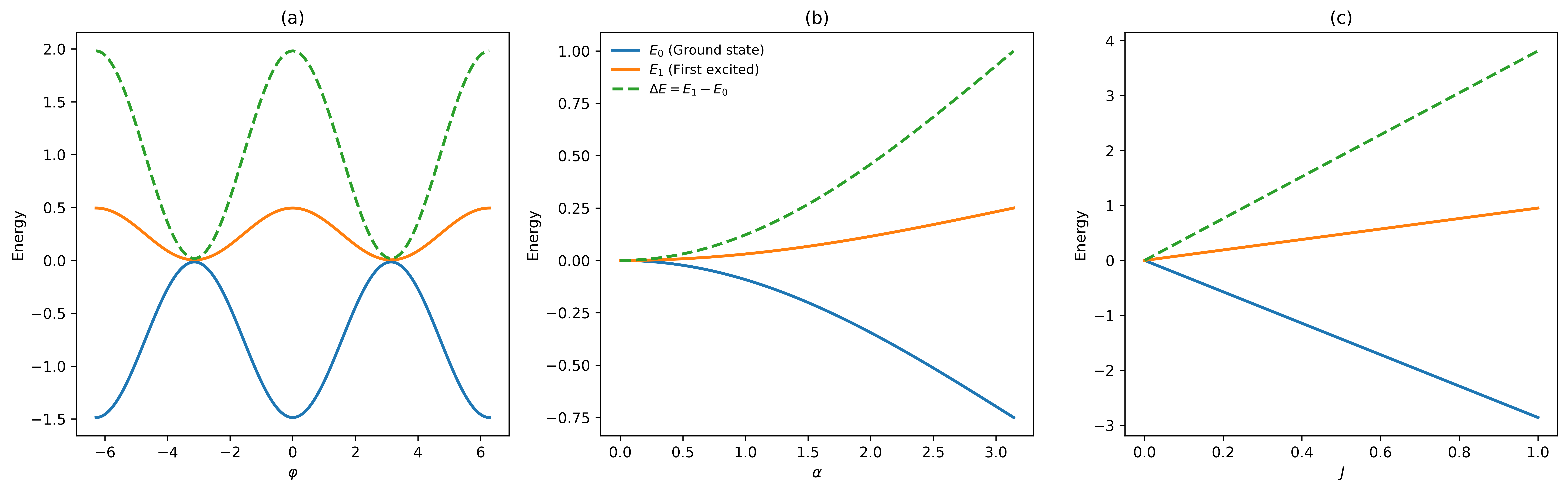} \caption{
Energy spectrum of the coupled Andreev spin-qubit system as a function of (a) the superconducting phase $\varphi$ for $\alpha=\pi/8$ and $J=0.5$, (b) the spin--orbit interaction strength $\alpha$ for $\varphi=\pi$ and $J=0.5$, and (c) the tunneling strength $J$ for $\alpha=\varphi=\pi$. The ground-state energy $E_0$ and the first-excited-state
energy $E_1$ are shown together with the corresponding energy gap $\Delta E=E_1-E_0$. The remaining excited-state levels overlap with $E_1$ and are therefore degenerate. The enhancement of the energy gap suppresses thermal population of the excited state, stabilizes the correlated ground state, and consequently leads to stronger thermal quantum correlations quantified by LQFI and LQU.}  \label{fig:gap} \end{figure*}

\subsection{Energy Gap and Thermal Stability}
To further clarify the physical origin of the behavior of quantum correlations, Fig.~\ref{fig:gap} presents the evolution of the ground-state energy, the first excited-state energy, and the corresponding energy gap $\Delta E=E_1-E_0,$ as functions of the superconducting phase difference $\varphi$, the spin--orbit interaction strength $\alpha$, and the tunneling amplitude $J$.
Since the thermal density matrix is weighted by the Boltzmann factors,
\begin{equation}\label{}
P_n=\frac{e^{-E_n/(k_BT)}}{Z},
\end{equation}
the quantity $\Delta E$ determines the thermal population of the excited state through
\begin{equation}\label{}
\frac{P_1}{P_0}=e^{-\Delta E/(k_BT)}.
\end{equation}
Consequently, the larger the energy gap, the less likely thermal excitations become, and the thermal state remains predominantly occupied by the ground state.

This observation provides a direct physical explanation for the behavior of LQFI and LQU discussed in the previous figures. In the parameter regions where the energy gap reaches its maximum values, thermal mixing is strongly suppressed and the coherence of the ground state is better preserved. Since the ground state possesses significant quantum correlations, both LQFI and LQU attain their largest values in these regions. Conversely, when the energy gap decreases, the first excited state becomes increasingly populated at finite temperatures, resulting in stronger thermal mixing and consequently weaker quantum correlations.

Figure~\ref{fig:gap}(a) shows that both the energy spectrum and the energy gap exhibit a periodic dependence on the superconducting phase difference. The energy gap is largest around $\varphi=0$ and $\varphi=\pm2\pi$, where the effective exchange couplings are strongest, while it reaches its minimum near $\varphi=\pm\pi$, where destructive interference suppresses the effective interactions. This behavior is fully consistent with the phase dependence of LQFI and LQU presented in Figs.~\ref{T_phi}, \ref{al_phi} and \ref{J_phi}.

Figure~\ref{fig:gap}(b) demonstrates that increasing the spin--orbit interaction strength generally enlarges the energy gap. The stronger spin--orbit coupling enhances the Dzyaloshinskii--Moriya interaction and modifies the effective exchange couplings, thereby increasing the stability of the entangled ground state against thermal excitations. This behavior agrees well with the quantum correlations depicted in Fig.~\ref{T_al}.

Finally, Fig.~\ref{fig:gap}(c) reveals that the tunneling amplitude $J$ increases the overall energy scale of the Hamiltonian. Consequently, the separation between the ground and excited states grows almost monotonically with increasing $J$, which explains why stronger tunneling leads to larger and more thermally robust quantum correlations. This behavior is in line with the quantum correlations illustrated in Fig.~\ref{T_J}.

\begin{figure*}[t] \centering \includegraphics[width=150mm]{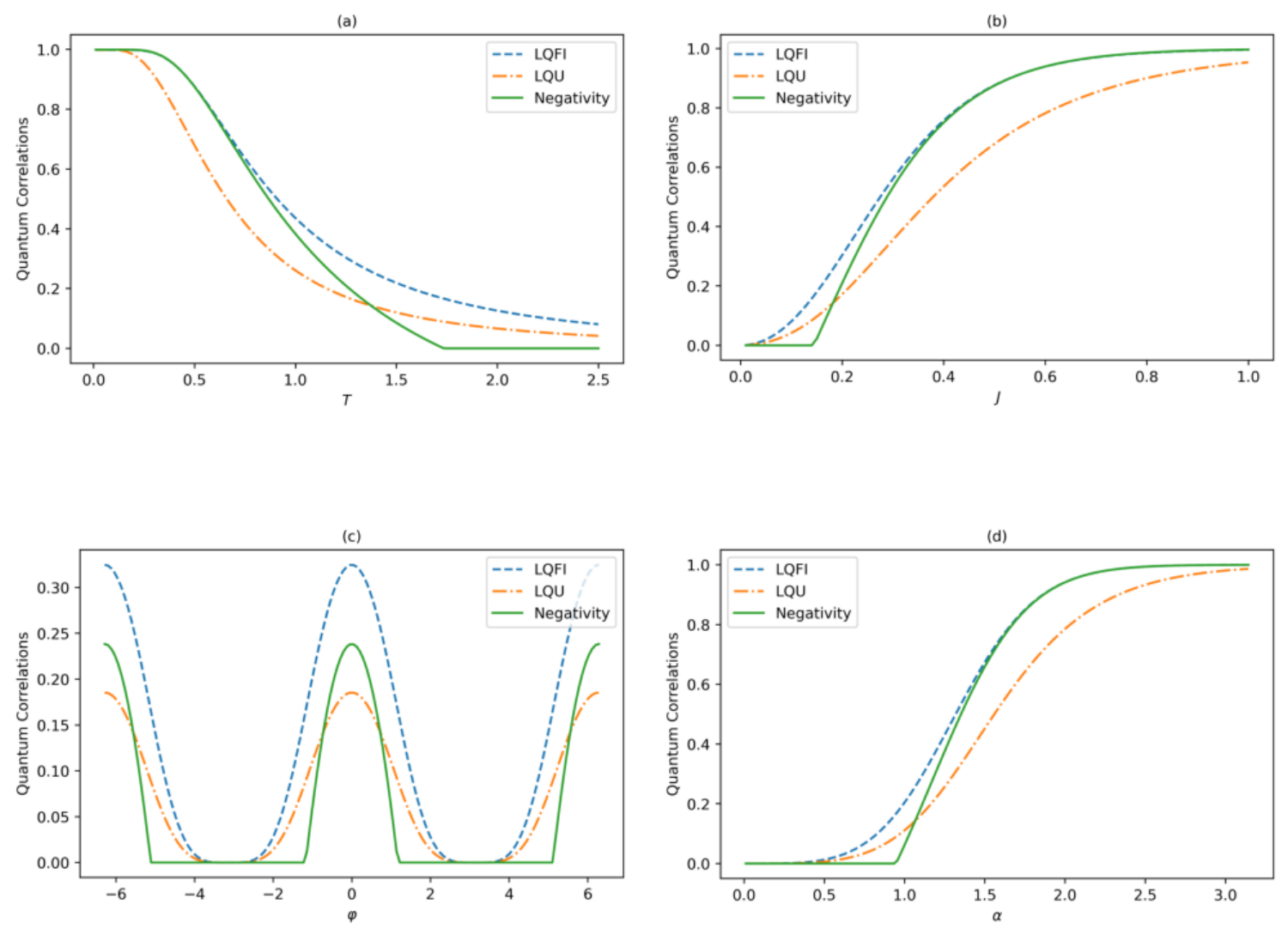} \caption{Comparison of LQFI, LQU and negativity as functions of (a) temperature $T$ for $J = 0.5$ and $\alpha = \varphi= \pi/8$, (b) coupling strength $J$ for $T = 0.5$ and $\alpha = \varphi= \pi/8$, (c) superconducting phase difference $\varphi$ for $J = 0.2$, $\alpha = \pi/8$ and $T = 0.5$, and (d) spin--orbit interaction strength $\alpha$ for $J=0.5$, $\varphi = \pi$ and $T = 0.1$. }  \label{fig:comparison} \end{figure*}
 
\subsection{Comparative Analysis of Quantum Correlations}
{
To further assess the behavior of the quantum correlations, we compare
the LQFI, LQU, and negativity in Fig.~\ref{fig:comparison}. In contrast to LQU
and LQFI, which quantify local quantum uncertainty and quantum
statistical distinguishability, respectively, negativity provides a
direct characterization of bipartite entanglement. For a bipartite
density matrix $\rho$, the negativity is defined as \cite{PhysRevA.65.032314}
\begin{equation}
\mathcal{N}(\rho)
=
\frac{\left\|\rho^{T_A}\right\|_1-1}{2},
\end{equation}
where $\rho^{T_A}$ denotes the partial transpose with respect to
subsystem $A$ and $\|\cdot\|_1$ is the trace norm. Equivalently, in
terms of the eigenvalues $\{\mu_k\}$ of the partially transposed density
matrix, it can be written as
\begin{equation}
\mathcal{N}(\rho)
=
\sum_{\mu_k<0}|\mu_k|.
\end{equation}
Thus, negativity vanishes when the partial transpose is positive
semidefinite and increases with the amount of negative eigenvalue
weight generated by the partial transpose. We note that for consistency in the numerical comparison, the negativity is multiplied by a factor of two to match the scale of LQFI and LQU.

As shown in Fig.~\ref{fig:comparison}(a), all three measures decrease
with increasing temperature. At low temperatures, the system is
predominantly populated by its low-energy correlated states, resulting
in relatively large values of LQFI, LQU, and negativity. With increasing
temperature, thermal mixing introduces contributions from excited
states and consequently suppresses the quantum correlations. An
important feature is that negativity reaches zero at a finite
temperature, whereas LQFI and LQU can remain finite beyond this
region. This behavior indicates that quantum correlations quantified
by LQFI and LQU may persist even after the entanglement, as detected by
negativity, has disappeared.

The dependence on the coupling strength $J$, shown in Fig.~\ref{fig:comparison}(b),
exhibits the opposite trend. Increasing $J$ enhances the effective
interaction between the two spin qubits and consequently increases
LQFI, LQU, and negativity. In particular, negativity develops from a
nearly vanishing value at weak coupling and approaches a large value
at sufficiently strong coupling. LQFI and LQU exhibit a qualitatively
similar increase, demonstrating that the strengthening of the
effective spin--spin interaction enhances both entanglement and more
general forms of nonclassical correlations.

The most characteristic behavior is observed as a function of the
superconducting phase difference $\varphi$, Fig.~\ref{fig:comparison}(c).
All three measures exhibit a pronounced periodic modulation. The
quantum correlations are enhanced around $\varphi=2n\pi$ with $n\in\{0,\pm 1\}$ and strongly suppressed around $\varphi=\pm \pi$. This behavior originates from
the phase dependence of the effective interaction coefficients in the
Andreev spin-qubit Hamiltonian. Importantly, LQFI, LQU, and negativity
follow the same overall phase-dependent pattern, although their
magnitudes and detailed profiles are different. This agreement
indicates that the phase-controlled modification of the effective
qubit interactions is reflected consistently in both entanglement and
the broader classes of quantum correlations quantified by LQFI and
LQU.

Finally, Fig.~\ref{fig:comparison}(d) shows the dependence of the three
measures on the spin--orbit interaction strength $\alpha$. Increasing
$\alpha$ generally enhances the quantum correlations. In particular,
the negativity increases from a nearly vanishing value at weak
spin--orbit interaction and approaches a saturated value for larger
$\alpha$. LQFI and LQU display a qualitatively similar enhancement,
although their growth rates and numerical values are not identical.
This behavior can be attributed to the spin--orbit-induced
Dzyaloshinskii--Moriya interaction, whose strength increases with
$\alpha$ and provides an additional mechanism for generating
correlations between the two qubits.

Overall, the comparison in Fig.~\ref{fig:comparison} reveals a strong
qualitative correspondence among LQFI, LQU, and negativity. All three
quantities respond consistently to the main control parameters of the
system: they are suppressed by thermal fluctuations and enhanced by
the effective coupling strength and spin--orbit interaction, while
exhibiting the same periodic phase dependence. Nevertheless, the three
measures are not equivalent. Negativity specifically quantifies
bipartite entanglement, whereas LQU and LQFI characterize broader
notions of local quantum uncertainty and quantum statistical
distinguishability. Consequently, the fact that LQFI or LQU remains
finite in parameter regimes where negativity vanishes indicates the
presence of nonclassical correlations beyond those captured by
entanglement alone. The similar qualitative trends, together with
their different quantitative behaviors, demonstrate that these
measures provide complementary descriptions of the quantum
correlations in the coupled Andreev spin-qubit system.}

\section{Conclusion}
In this work, we have investigated thermal quantum correlations in a system of two coupled superconducting spin qubits described by an effective Andreev spin-qubit Hamiltonian in the presence of spin--orbit interaction. By employing LQFI and LQU as quantum-correlation measures, we systematically analyzed the effects of temperature, superconducting phase difference, tunneling amplitude, and spin--orbit interaction on the nonclassical properties of the thermal state.

Our results show that thermal fluctuations play a destructive role by increasing the mixedness of the density matrix and consequently suppressing quantum correlations. In contrast, increasing the tunneling amplitude and the spin--orbit interaction significantly enhances the robustness of quantum correlations against thermal effects through the strengthening of the effective exchange interactions.

The superconducting phase difference was found to be one of the key control parameters of the system. Owing to its explicit appearance in the effective coupling coefficients, it periodically modulates the exchange interactions and produces alternating constructive and destructive interference regimes. Consequently, quantum correlations are generally enhanced near $\varphi=0$, where the effective couplings are reinforced, while they are strongly suppressed near $\varphi=\pi$, where the effective exchange interactions become considerably weakened.

A particularly important result of the present work is the analysis of the energy spectrum of the effective Hamiltonian. We showed that the energy gap between the ground state and the first excited state provides a useful physical interpretation of the thermal behavior of LQFI and LQU. A larger energy gap suppresses the thermal population of the excited states, stabilizes the correlated ground state, and therefore preserves quantum coherence and quantum correlations over a wider temperature range. This provides a clear microscopic interpretation of the numerical results obtained throughout the paper.

Another important finding is that LQFI remains consistently larger than or equal to LQU over the entire investigated parameter space, satisfying LQFI $\geq$ LQU, in agreement with the theoretical hierarchy between these quantities. This persistent ordering reflects the different sensitivities of the two measures to local quantum fluctuations and highlights the broader response captured by LQFI. In addition, the comparison with negativity provides a complementary assessment of entanglement, allowing the behavior of LQFI and LQU to be contrasted with that of a standard entanglement measure under the same physical conditions.

Overall, our results demonstrate that superconducting phase engineering, tunneling processes, and spin--orbit interaction provide efficient and experimentally accessible tools for controlling thermal quantum correlations in Andreev spin qubits. Beyond the specific model studied here, the physical mechanism revealed through the interplay between effective exchange interactions and the energy spectrum may also be relevant to other hybrid superconducting-spin architectures. We therefore expect that the present results will be useful for the development of robust solid-state quantum-information devices, quantum sensing and metrology protocols, and scalable superconducting quantum technologies.

\section*{ORCID iDs}
Mohammad Reza Pourkarimi \href{https://orcid.org}{https://orcid.org/0000-0002-8554-1396}\\
Mina shiri  \href{https://orcid.org}{https://orcid.org/0009-0004-9896-1121}\\
Mehdi Monemi \href{https://orcid.org}{https://orcid.org/0000-0003-0258-5156}\\
Vadim Ohanyan \href{https://orcid.org}{https://orcid.org/0000-0002-7810-7321}

\bibliographystyle{naturemag}

\bibliography{refre3}

\end{document}